\documentclass[12pt,showpacs,showkeys,amsmath,amssymb]{revtex4}
\usepackage{amsmath,amsfonts,amsthm,amscd,amssymb,latexsym}
\usepackage{bm}
\usepackage{dcolumn}
\usepackage{graphicx}
\usepackage{epstopdf}
\usepackage{color}
\usepackage{epsf}
\usepackage{epsfig}
\usepackage{graphicx, epic, eepic, color}

\newcommand{\beq}{\begin{equation}}
\newcommand{\eeq}{\end{equation}}

\def\@{\partial_}

\def\negenspace{\kern-1.1em}

\def\sqr#1#2{{\vcenter{\hrule height.#2pt\hbox{\vrule width.#2pt
height#1pt \kern#1pt \vrule width.#2pt}\hrule height.#2pt}}}

\begin{document}

\title{Nonlocal Gravity: Fundamental Tetrads and Constitutive Relations}

\author{Bahram \surname{Mashhoon}$^{1,2}$}
\email{mashhoonb@missouri.edu}

\affiliation{$^1$Department of Physics and Astronomy, University of Missouri, Columbia, Missouri 65211, USA\\
$^2$School of Astronomy, Institute for Research in Fundamental
Sciences (IPM), P. O. Box 19395-5531, Tehran, Iran\\
}

\date{\today}

\begin{abstract}
Nonlocal gravity (NLG) is a classical nonlocal generalization of Einstein's theory of gravitation based on a certain analogy with the nonlocal electrodynamics of media. The history dependence enters NLG through a constitutive relation involving a causal kernel that should ultimately be determined via observational data. The purpose of this paper is to reformulate nonlocal gravity such that the nonlocal aspect of the constitutive relation directly connects measurable quantities as in the nonlocal electrodynamics of media. The resulting constitutive relation turns out to coincide with the recent suggestion of Puetzfeld, Obukhov, and Hehl~\cite{Puetzfeld:2019wwo}. With the new constitutive relation of NLG, it is possible to show that de Sitter spacetime is not a solution of NLG. 
\\
$~~~~~~~~~~~~~~~~$ \emph{Dedicated to Friedrich W. Hehl on his eighty-fifth birthday.}
\end{abstract}


\maketitle

\section{Introduction}

In the theory of relativity,  gravitation has to do with the deviation of the spacetime manifold from flat Minkowski spacetime. The fundamental microphysical laws of physics have been formulated with respect to the ideal inertial observers that are all at rest in a global inertial frame in Minkowski spacetime with Cartesian coordinates $X^{ \mu} = (cT, \mathbf{X})$ and corresponding metric
\begin{equation}\label{I1}
dS^2 = \eta_{ \alpha  \beta} dX^{ \alpha}dX^{ \beta}\,,
\end{equation}
where the Minkowski metric tensor $\eta_{\mu \nu}$ is given by diag$(-1,1,1,1)$ and we use units such that $c= 1$, unless specified otherwise.  Moreover, in our convention, Greek indices run from 0 to 3, while Latin indices run from 1 to 3.
The ideal inertial observers carry orthonormal tetrads $^{M}e^{\mu}{}_{\hat \alpha} = \delta^\mu_\alpha$ that consist of unit vectors that point along the Cartesian coordinate axes and are therefore globally parallel. We use hatted indices to enumerate tetrad axes in the local tangent space, while indices without hats are ordinary spacetime indices.  Minkowski spacetime contains an equivalence class of all such parallel frame fields that are related to each other by constant elements of the global Lorentz group. 

Inertial observers may choose any smooth admissible system of curvilinear coordinates $x^\mu = x^\mu(X^{ \alpha})$ in Minkowski spacetime with the corresponding metric 
\begin{equation}\label{I2}
dS^2 = \gamma_{\mu \nu} dx^\mu\,dx^\nu\,.
\end{equation}
The associated fundamental globally parallel tetrad frame field is now given by $^{M}E^{\mu}{}_{\hat \alpha} = \partial x^\mu/\partial X^{ \alpha}$ and the orthonormality condition takes the form  
\begin{equation}\label{I3}
 \eta_{\hat \alpha \hat \beta}  =\gamma_{\mu \nu}\,^{M}E^{\mu}{}_{\hat \alpha}\,^{M}E^{\nu}{}_{\hat \beta}\,.
\end{equation}

Nonlocal gravity (NLG) is a classical history-dependent generalization of Einstein's general relativity (GR) theory patterned after the nonlocal electrodynamics of media. Indeed, the theory involves a certain average of the gravitational field over past events.  The purpose of the 16 partial integro-differential equations that constitute the field equation of NLG is to find the 16 components of $e^{\mu}{}_{\hat \alpha}$, the fundamental tetrad frame field of the theory. The fundamental tetrads are adapted to the fundamental observers of NLG theory. The preferred tetrads are orthonormal, namely, 
\begin{equation}\label{I4}
 \eta_{\hat \alpha \hat \beta}  = g_{\mu \nu}e^{\mu}{}_{\hat \alpha}e^{\nu}{}_{\hat \beta}\,, \qquad  g^{\mu \nu}  = \eta^{\hat \alpha \hat \beta}e^{\mu}{}_{\hat \alpha}e^{\nu}{}_{\hat \beta}\,,
\end{equation}
where $g_{\mu \nu}(x)$ is the spacetime metric in arbitrary smooth admissible coordinates,
\begin{equation}\label{I5}
ds^2 = g_{\mu \nu}(x) dx^\mu\,dx^\nu\,.
\end{equation}  
In the absence of the gravitational field, the fundamental observers reduce to the ideal global inertial observers at rest in an inertial frame in Minkowski spacetime with $e^{\mu}{}_{\hat \alpha}(x) =\, ^{M}E^{\mu}{}_{\hat \alpha}(x)$. The description of NLG requires an extended GR framework. A comprehensive account of NLG is contained in Ref.~\cite{BMB}. 

As will be explained in some detail in the next section, the introduction of history dependence in NLG is based on a certain analogy with Maxwell's electrodynamics of media. Consider Maxwell's equations in a medium in Minkowski spacetime. The electromagnetic field strength is given by $(\mathbf{E}, \mathbf{B}) \mapsto F_{\mu\nu}$ such that in an inertial reference frame, source-free Maxwell's equations imply
\begin{equation}\label{I6}
F_{\mu \nu} = \partial_{ \mu}A_{ \nu} -\partial_{ \nu}A_{ \mu}\,,
\end{equation}
where $A_\mu$ is the vector potential. The medium, in the presence of $F_{\mu \nu}$, responds via its polarizability and magnetizability resulting in net electromagnetic field excitations $(\mathbf{D}, \mathbf{H}) \mapsto H_{\mu\nu}$, such that 
\begin{equation}\label{I7}
\partial_{\nu}H^{\mu \nu} = \frac{4 \pi}{c} J^{\mu}\,,
\end{equation}
where $J^\mu$ is the current 4-vector associated with \emph{free} electric charges. For the fundamental ideal inertial observers, the field variables are the same as the observed quantities in Equations~\eqref{I6} and~\eqref{I7}, namely, field projections on the fundamental tetrads. That is, $H_{\hat \alpha \hat \beta} = H_{\mu \nu}\,^{M}e^{\mu}{}_{\hat \alpha}\,^{M}e^{\nu}{}_{\hat \beta} = H_{\alpha \beta}$ and $F_{\hat \alpha \hat \beta} = F_{\mu \nu}\,^{M}e^{\mu}{}_{\hat \alpha}\,^{M}e^{\nu}{}_{\hat \beta} = F_{\alpha \beta}$, since $^{M}e^{\mu}{}_{\hat \alpha} = \delta^\mu_\alpha$. The constitutive relation, which connects \emph{measured} quantities $H_{\hat \alpha \hat \beta}$ and $F_{\hat \alpha \hat \beta}$, is characteristic of the background medium. For instance, for a medium where the relation is local and linear, 
\begin{equation}\label{I8}
H_{\hat \alpha \hat \beta} = \frac{1}{2}\,\chi_{\hat \alpha \hat \beta}{}^{\hat \mu \hat \nu}\,F_{\hat \mu \hat \nu}\,,
\end{equation}
where $\chi_{\hat \alpha \hat \beta}{}^{\hat \mu \hat \nu}$ is the electromagnetic constitutive tensor that is antisymmetric in its first and last two indices.  

The electromagnetic properties of material media, especially magnetic materials, generally exhibit history dependence (``hysteresis"). The causal connection between the input (
$F_{\hat \alpha \hat \beta}$) and the output ($H_{\hat \alpha \hat \beta}$) could be nonlinear; however, for the sake of simplicity, we assume linearity throughout this work~\cite{L+L, Jack, HeOb}.  

The main purpose of this paper is to reformulate the constitutive relation of NLG theory in complete correspondence with the electrodynamics of media described above. That is, the new constitutive relation will be formulated in such a way that it corresponds to measurable quantities as determined by the preferred observers of the theory and their adapted fundamental tetrad frame field $e^{\mu}{}_{\hat \alpha}(x)$.   

\section{GR and Teleparallelism}

To describe nonlocal gravity as an extension of GR, we need a framework that involves the Levi-Civita connection as well as the Weitzenb\"ock  connection. They are both compatible with the spacetime metric tensor $g_{\mu \nu}(x)$. As in GR~\cite{Einstein}, free test particles and light rays follow timelike and null geodesics, respectively. 
 The symmetric Levi-Civita connection is given by
\begin{equation}\label{G1}
{^0}\Gamma^\mu_{\alpha \beta}= \frac{1}{2} g^{\mu \nu} (g_{\nu \alpha,\beta}+g_{\nu \beta,\alpha}-g_{\alpha \beta,\nu})\,
\end{equation}
and its associated Riemann curvature tensor is
\begin{equation}\label{G2}
^{0}R^{\alpha}{}_{\mu \beta \nu}=\partial_{\beta}\, {^0} \Gamma^\alpha_{\nu \mu} -\partial_{\nu}\, {^0} \Gamma^\alpha_{\beta \mu}+\,^{0}\Gamma^{\alpha}_{\beta \gamma}\, ^{0}\Gamma^{\gamma}_{\nu \mu}-\,^{0}\Gamma^\alpha_{\nu \gamma}\, ^{0}\Gamma^\gamma_{\beta \mu}\,.
\end{equation}
 A left superscript ``0"  is employed to refer to all geometric quantities related to the Levi-Civita connection.  Einstein's gravitational field equation is given by~\cite{Einstein}
\begin{equation}\label{G3}
{^0}G_{\mu \nu} + \Lambda\, g_{\mu \nu}=\kappa\,T_{\mu \nu}\,, 
 \end{equation}
 where ${^0}G_{\mu \nu}$ is the Einstein tensor  
\begin{equation}\label{G4}
 {^0}G_{\mu \nu} := {^0}R_{\mu \nu}-\frac{1}{2} g_{\mu \nu}\,{^0}R\,.
 \end{equation} 
Here, $T_{\mu \nu}$ is the symmetric energy-momentum tensor of matter, $\Lambda$ is the cosmological constant and $\kappa:=8 \pi G/c^4$.  The trace of the Riemann tensor, 
${^0}R^{\alpha}{}_{\mu \alpha \nu} = {^0}R_{\mu \nu}$,  is the Ricci tensor and its trace, $g^{\mu \nu}\,{^0}R_{\mu \nu} = {^0}R$, is the scalar curvature. In GR, Einstein's equation with $\Lambda = 0$ reduces to  Poisson's equation of Newtonian gravitation in the correspondence limit (where we formally let $c \to \infty$).

Next, consider a smooth orthonormal frame field $e^{\mu}{}_{\hat \alpha}(x)$ adapted to a congruence of preferred observers in spacetime. We use the frame field to define  the \emph{Weitzenb\"ock} connection~\cite{We}
\begin{equation}\label{G5}
\Gamma^\mu_{\alpha \beta}=e^\mu{}_{\hat{\rho}}~\partial_\alpha\,e_\beta{}^{\hat{\rho}}\,.
\end{equation}
This connection is nonsymmetric and curvature free. The Weitzenb\"ock covariant derivative of the preferred tetrad frame vanishes, i.e., 
\begin{equation}\label{G6}
\nabla_\nu\,e_\mu{}^{\hat{\alpha}}=0\,,
\end{equation}
which means that the fundamental frame field is globally parallel. The existence of a global set of parallel frame fields renders the spacetime manifold parallelizable. In this framework of teleparallelism~\cite{Maluf:2011kf, Aldrovandi:2013wha, Maluf:2013gaa}, two distant vectors, tensors, etc., are defined to be parallel to each other if they have the same components with respect to their local preferred tetrad frames. Furthermore, Equation~\eqref{G6} implies that  the Weitzenb\"ock connection is compatible with the spacetime metric; that is, 
\begin{equation}\label{G7}
\nabla_\gamma\,g_{\alpha \beta}=0\,, \qquad  g_{\alpha \beta , \gamma}= \Gamma^\mu_{\gamma \alpha}\, g_{\mu \beta} + \Gamma^\mu_{\gamma \beta}\, g_{\mu \alpha}\,.
\end{equation}

 The difference between two connections on a manifold is a tensor.  For the Weitzenb\"ock connection,  we define the \emph{torsion} tensor by
\begin{equation}\label{G8}
 C_{\mu \nu}{}^{\alpha}:=\Gamma^{\alpha}_{\mu \nu}-\Gamma^{\alpha}_{\nu \mu}=e^\alpha{}_{\hat{\beta}}\Big(\partial_{\mu}e_{\nu}{}^{\hat{\beta}}-\partial_{\nu}e_{\mu}{}^{\hat{\beta}}\Big)\,
\end{equation}
 and the \emph{contorsion} tensor by
\begin{equation}\label{G9}
K_{\mu \nu}{}^\alpha= {^0} \Gamma^\alpha_{\mu \nu} - \Gamma^\alpha_{\mu \nu}\,.
\end{equation}
It can be shown that~\cite{BMB}
\begin{equation}\label{G10}
K_{\mu \nu}{}^\alpha = \frac{1}{2}\, g^{\alpha \beta} (C_{\mu \beta \nu}+C_{\nu \beta \mu}-C_{\mu \nu \beta})\,.
\end{equation}
 The contorsion tensor $K_{\alpha \beta \gamma}$ is antisymmetric in its last two indices, while the torsion tensor $C_{\alpha \beta \gamma}$ is antisymmetric in its first two indices. It proves interesting to introduce an auxiliary torsion tensor via
\begin{equation}\label{G11}
\mathfrak{C}_{\alpha \beta \gamma} :=C_\alpha\, g_{\beta \gamma} - C_\beta \,g_{\alpha \gamma}+K_{\gamma \alpha \beta}\,,
\end{equation} 
where $C_\mu$ is the torsion vector $C_\mu :=C^{\alpha}{}_{\mu \alpha} = - C_{\mu}{}^{\alpha}{}_{\alpha}$.  

The field equation of NLG can be expressed as
\begin{equation}\label{G12}
\frac{\partial}{\partial x^\nu}\,\mathcal{H}^{\mu \nu}{}_{\hat{\alpha}}+\frac{\sqrt{-g}}{\kappa}\,\Lambda\,e^\mu{}_{\hat{\alpha}} =\sqrt{-g}\,(T_{\hat{\alpha}}{}^\mu + \mathcal{T}_{\hat{\alpha}}{}^\mu)\,,
\end{equation}
where $g:=\det(g_{\mu \nu})$, $\sqrt{-g}=\det(e_{\mu}{}^{\hat{\alpha}})$ and
\begin{equation}\label{G13}
\mathcal{H}_{\mu \nu \rho}:= \frac{\sqrt{-g}}{\kappa}\,(\mathfrak{C}_{\mu \nu \rho} + N_{\mu \nu \rho})\,.
\end{equation}
Here, $N_{\mu \nu \rho} = - N_{\nu \mu \rho}$  is the nonlocality tensor. The antisymmetry of $\mathcal{H}_{\mu \nu \rho}$ in its first two indices implies
\begin{equation}\label{G14}
\frac{\partial}{\partial x^\mu}\,\Big[\sqrt{-g}\,(T_{\hat{\alpha}}{}^\mu + \mathcal{T}_{\hat{\alpha}}{}^\mu -\frac{\Lambda}{\kappa}\,e^\mu{}_{\hat{\alpha}})\Big]=0\,,
 \end{equation}
which expresses the law of conservation of total energy-momentum tensor. That is, $T_{\mu \nu}$ is, as before, the symmetric energy-momentum tensor of matter, while $\mathcal{T}_{\mu \nu}$ is the traceless energy-momentum tensor of the gravitational field in nonlocal gravity, 
\begin{equation}\label{G15}
\sqrt{-g}\,\mathcal{T}_{\mu \nu} :=C_{\mu \rho \sigma}\, \mathcal{H}_{\nu}{}^{\rho \sigma}-\tfrac{1}{4}\, g_{\mu \nu}\,C_{\rho \sigma \delta}\,\mathcal{H}^{\rho \sigma \delta}\,.
\end{equation}

Before we specify the nonlocality tensor $N_{\mu \nu \rho}$, it is necessary to point out that in the absence of this tensor, $N_{\mu \nu \rho} = 0$, the theory described by Equations~\eqref{G12}--\eqref{G15} is indeed equivalent to Einstein's GR.

\subsection{Teleparallel Equivalent of General Relativity (TEGR)}

We can start from ${^0} \Gamma^\alpha_{\mu \nu} = \Gamma^\alpha_{\mu \nu} + K_{\mu \nu}{}^\alpha$ and express GR in terms of the torsion tensor. The Einstein tensor can be written as~\cite{BMB} 
\begin{eqnarray}\label{G16}
 {^0}G_{\mu \nu}=\frac{\kappa}{\sqrt{-g}}\Big[e_\mu{}^{\hat{\gamma}}\,g_{\nu \alpha}\, \frac{\partial}{\partial x^\beta}\,\mathfrak{H}^{\alpha \beta}{}_{\hat{\gamma}}
-\Big(C_{\mu}{}^{\rho \sigma}\,\mathfrak{H}_{\nu \rho \sigma} -\tfrac{1}{4}\,g_{\mu \nu}\,C^{\alpha \beta \gamma}\,\mathfrak{H}_{\alpha \beta \gamma}\Big) \Big]\,,
\end{eqnarray}
where $\mathfrak{H}_{\mu \nu \rho}$ is defined by 
\begin{equation}\label{G17}
\mathfrak{H}_{\mu \nu \rho}:= \frac{\sqrt{-g}}{\kappa}\,\mathfrak{C}_{\mu \nu \rho}\,.
\end{equation}
Einstein's field equation~\eqref{G3} then takes the form
\begin{equation}\label{G18}
 \frac{\partial}{\partial x^\nu}\,\mathfrak{H}^{\mu \nu}{}_{\hat{\alpha}}+\frac{\sqrt{-g}}{\kappa}\,\Lambda\,e^\mu{}_{\hat{\alpha}} =\sqrt{-g}\,(T_{\hat{\alpha}}{}^\mu + \mathbb{T}_{\hat{\alpha}}{}^\mu)\,,
\end{equation}
where $\mathbb{T}_{\mu \nu}$ is the trace-free energy-momentum tensor of the gravitational field in TEGR, namely, 
\begin{equation}\label{G19}
\kappa\,\mathbb{T}_{\mu \nu} :=C_{\mu \rho \sigma}\, \mathfrak{C}_{\nu}{}^{\rho \sigma}-\tfrac{1}{4}\, g_{\mu \nu}\,C_{\rho \sigma \delta}\,\mathfrak{C}^{\rho \sigma \delta}\,.
\end{equation}
The total energy-momentum tensor is conserved,
\begin{equation}\label{G20}
\frac{\partial}{\partial x^\mu}\,\Big[\sqrt{-g}\,(T_{\hat{\alpha}}{}^\mu + \mathbb{T}_{\hat{\alpha}}{}^\mu -\frac{\Lambda}{\kappa}\,e^\mu{}_{\hat{\alpha}})\Big]=0\,.
 \end{equation}

This form of Einstein's theory bears a certain resemblance to Maxwell's electrodynamics. The spacetime torsion~\eqref{G8} is analogous to the electromagnetic field strength. Indeed, for each $\hat \alpha$ in
\begin{equation}\label{G21}
 C_{\mu \nu}{}^{\hat \alpha}=\partial_{\mu}e_{\nu}{}^{\hat{\alpha}}-\partial_{\nu}e_{\mu}{}^{\hat{\alpha}}\,,
\end{equation}
we have an expression relating $C_{\mu \nu}{}^{\hat \alpha}$ with $e_{\mu}{}^{\hat \alpha}$ that is reminiscent of the connection between the electromagnetic field tensor $F_{\mu \nu}$ and the vector potential $A_\mu$. Furthermore, the auxiliary torsion field $\mathfrak{H}^{\mu \nu}{}_{\hat{\alpha}}$ in Equation~\eqref{G18} is analogous to the electromagnetic excitation in Equation~\eqref{I7}. Finally, we can regard Equation~\eqref{G17}, namely,
\begin{equation}\label{G22}
\mathfrak{H}_{\alpha \beta \gamma} = \frac{\sqrt{-g}}{\kappa}\,\left[\tfrac{1}{2}(C_{\gamma \alpha \beta} + C_{\alpha \beta \gamma} -C_{\gamma \alpha \beta}) +C_\alpha\, g_{\beta \gamma} - C_\beta \,g_{\alpha \gamma}\right]\,,
\end{equation} 
as the local constitutive relation of TEGR, since it connects $\mathfrak{H}_{\alpha \beta \gamma}$ to $C_{\alpha \beta \gamma}$. 

The source of the analogy with electrodynamics has to do with the fact that TEGR is the  gauge theory of the 4-parameter Abelian group of spacetime translations~\cite{Cho}. Therefore, TEGR, though nonlinear, is formally analogous to electrodynamics and can be rendered nonlocal via history-dependent constitutive relations as in the nonlocal electrodynamics of media. These considerations led F. W. Hehl to suggest that GR could be made nonlocal in this way and the idea was subsequently worked out in Refs.~\cite{Hehl:2008eu, Hehl:2009es}.  For further discussion of these ideas, see~\cite{BMB, BlHe, Itin:2018dru, Mashhoon:2019jkq, Hehl:2020hhp,  Puetzfeld:2020tkp, Obukhov:2020uan}.

To introduce history dependence as in the nonlocal electrodynamics of media, we modify the constitutive relation of TEGR while keeping the gravitational field equation intact; equivalently, NLG reduces to TEGR when $N_{\mu \nu \rho} = 0$. This connection between NLG and TEGR makes it possible to find the nonlocally modified Einstein's field equation. To this end, we use Equations~\eqref{G13} and~\eqref{G17} to write
\begin{equation}\label{G23}
\mathfrak{H}_{\mu \nu \rho}:= \mathcal{H}_{\mu \nu \rho} - \frac{\sqrt{-g}}{\kappa}\,N_{\mu \nu \rho}\,.
\end{equation}
Substituting this relation in Equation~\eqref{G16}  and using the field Equation~\eqref{G12} of NLG , we get the nonlocal GR field equation
 \begin{equation}\label{G24}
^{0}G_{\mu \nu}  + \Lambda g_{\mu \nu} = \kappa T_{\mu \nu} - \mathcal{N}_{\mu \nu} +  Q_{\mu \nu}\,.
\end{equation}
Here,  $\mathcal{N}_{\mu \nu}$  is a nonlocal tensor given by
\begin{equation}\label{G25}
\mathcal{N}_{\mu \nu} = g_{\nu \alpha} e_\mu{}^{\hat{\gamma}} \frac{1}{\sqrt{-g}} \frac{\partial}{\partial x^\beta}\,(\sqrt{-g}N^{\alpha \beta}{}_{\hat{\gamma}})\,
\end{equation} 
and $Q_{\mu \nu} := \kappa (\mathcal{T}_{\mu \nu} - \mathbb{T}_{\mu \nu})$ is traceless, i.e.,  
\begin{equation}\label{G26}
Q_{\mu \nu} = C_{\mu \rho \sigma} N_{\nu}{}^{\rho \sigma}-\tfrac{1}{4}\, g_{\mu \nu}\,C_{ \delta \rho \sigma}N^{\delta \rho \sigma}\,.
\end{equation} 

It is natural to split the nonlocal GR field equation into its symmetric and antisymmetric parts; that is, 
 \begin{equation}\label{G27}
^{0}G_{\mu \nu}  + \Lambda g_{\mu \nu} = \kappa T_{\mu \nu} - \mathcal{N}_{(\mu \nu)} +  Q_{(\mu \nu)}\,
\end{equation}
and
 \begin{equation}\label{G28}
\mathcal{N}_{[\mu \nu]} =  Q_{[\mu \nu]}\,.
\end{equation}
Of the 16 components of the fundamental tetrad $e^\mu{}_{\hat{\alpha}}$,  10 fix the components of the metric tensor $g_{\mu \nu}$ via the orthonormality condition~\eqref{I4}, while the other 6 are local Lorentz degrees of freedom (i.e., boosts and rotations). Similarly, as illustrated by Equations~\eqref{G27} and~\eqref{G28}, the 16 field equations of NLG for the 16 components of the fundamental tetrad $e^\mu{}_{\hat{\alpha}}$ naturally split into 10 nonlocally modified equations of GR  plus 6 integral constraint equations for the nonlocality tensor $N_{\mu \nu \rho}$. These constraints disappear in the Newtonian regime of NLG, while the nonlocal modification of GR has the interpretation of effective dark matter~\cite{BMB, Rahvar:2014yta, Chicone:2015coa, Roshan:2021ljs, Roshan:2022zov, Roshan:2022ypk}.

It remains to specify the exact nonlocal connection between $N_{\mu \nu \rho}$ and $C_{\mu \nu \rho}$. 

\subsection{Old Nonlocal Constitutive Relation}

In NLG, we have assumed that~\cite{BMB}
\begin{equation}\label{N1}
N_{\mu \nu \rho}(x)  = - \int \Omega_{\mu \mu'} \Omega_{\nu \nu'} \Omega_{\rho \rho'}\, {\cal K}(x, x')\,X^{\mu' \nu' \rho'}(x') \sqrt{-g(x')}\, d^4x' \,,
\end{equation} 
where $\Omega(x, x')$ is Synge's \emph{world function}~\cite{Sy}, $\mathcal{K}$ is the  \emph{causal} scalar kernel of the nonlocal theory and  $X_{\mu \nu \rho}(x)$ is a tensor that is antisymmetric in its first two indices and is given by
\begin{equation}\label{N2}
X_{\mu \nu \rho}= \mathfrak{C}_{\mu \nu \rho}+ \check{p}\,(\check{C}_\mu\, g_{\nu \rho}-\check{C}_\nu\, g_{\mu \rho})\,.
\end{equation}
Here, $\check{p}\ne 0$ is a constant dimensionless parameter and  $\check{C}^\mu$ is the torsion pseudovector defined via the Levi-Civita tensor $E_{\alpha \beta \gamma \delta}$ by
\begin{equation}\label{N3}
\check{C}_\mu :=\frac{1}{3!} C^{\alpha \beta \gamma}\,E_{\alpha \beta \gamma \mu}\,.
\end{equation}
Let us note that the relationship between $X_{\mu \nu \rho}$ and $C_{\alpha \beta \gamma}$ in Equation~\eqref{N2} is local and linear and is of the general form
\begin{equation}\label{N4}
X_{\mu \nu \rho} = \frac{1}{2} \,\chi_{\mu \nu \rho}{}^{\alpha \beta \gamma}\,C_{\alpha \beta \gamma}\,.
\end{equation}
Gravitational constitutive tensors $\chi_{\mu \nu \rho}{}^{\alpha \beta \gamma}$ have been thoroughly studied and classified in Ref.~\cite{Itin:2018dru}.

At first sight, the constitutive relation~\eqref{N1}, which involves a spacetime average of the gravitational field (i.e., torsion tensor) over past events via a causal constitutive kernel,  appears natural and simple, since $\Omega_{\mu \mu'} (x, x') = \Omega_{\mu' \mu} (x, x')$ is a bitensor that is dimensionless and has a natural coincidence limit in terms of the spacetime metric tensor, namely, for $x' \to x$, $\Omega_{\mu \mu'} (x, x')  \to - g_{\mu \mu'}(x)$. In practice, however, this bitensor has a complicated mathematical structure~\cite{Puetzfeld:2019wwo}. Furthermore, it has not been possible to find a nontrivial solution of NLG theory. Indeed, 
it is important to point out that the only known exact solution of NLG is the trivial solution; that is, we recover Minkowski spacetime in the absence of the gravitational field.  The structure of Equation~\eqref{N1} appears to be partly responsible for the fact that no exact nontrivial solution of NLG is known. 

The known observational implications of NLG are all based on the linearized form of Equation~\eqref{N1}; that is, to first order in the deviation from Minkowski spacetime, the implications of linearized NLG have been extensively studied~\cite{BMB, Rahvar:2014yta, Chicone:2015coa, Roshan:2021ljs, Roshan:2022zov}. In searching for a replacement for Equation~\eqref{N1}, we must make sure that this constitutive relation is preserved at the linear order.

\subsection{New Nonlocal Constitutive Relation}

In nonlocal electrodynamics, the components of $H_{\mu \nu}$, as measured by the fundamental inertial observers in Minkowski spacetime, namely, $H_{\hat \alpha \hat \beta}$, are connected to the corresponding measured components of $F_{\mu \nu}$, namely, $F_{\hat \alpha \hat \beta}$, via the constitutive relation of the theory. That is, the input of the constitutive relation is $F_{\hat \mu \hat \nu}$ and the output is $H_{\hat \mu \hat \nu}$.
\emph{The analogy with the nonlocal electrodynamics of media suggests that the components of $N_{\mu \nu \rho}$, as measured by the fundamental observers of the theory with adapted tetrads 
$e^\mu{}_{\hat{\alpha}}$, must be physically related to the corresponding measured components of $X_{\mu \nu \rho}$; that is, we must replace Equation~\eqref{N1} with 
\begin{equation}\label{N4}
N_{\hat \mu \hat \nu \hat \rho}(x) =  \int  \mathcal{K}(x, x')\,X_{\hat \mu  \hat \nu  \hat \rho }(x') \sqrt{-g(x')}\, d^4x' \,,
\end{equation} 
where 
\begin{equation}\label{N5}
X_{\hat \mu \hat \nu \hat \rho}= \mathfrak{C}_{\hat \mu \hat \nu \hat \rho}+ \check{p}\,(\check{C}_{\hat \mu}\, \eta_{\hat \nu \hat \rho}-\check{C}_{\hat \nu}\, \eta_{\hat \mu \hat \rho})\,.
\end{equation}
}
In this way, the new constitutive relation directly refers to \emph{scalar} gravitational field quantities.  

The new constitutive relation~\eqref{N4} coincides with the simple form of the nonlocal constitutive relation previously suggested in Ref.~\cite{Puetzfeld:2019wwo}, where the  bitensor 
$-\Omega_{\mu \mu'}(x, x')$  is replaced by the parallel propagator $e_{\mu}{}^{\hat{\alpha}}(x)\,e_{\mu' {}\hat{\alpha}}(x')$ for the sake of simplicity. This is possible within the framework of teleparallelism. That is, independently of the issue of measurability of field quantities that appear in the new constitutive relation,  the fundamental  orthonormal frame field $e^\mu{}_{\hat{\alpha}}$ is parallel throughout spacetime.    At the linear order, the new ansatz coincides with the old one, as already noted in Ref.~\cite{Puetzfeld:2019wwo} as well. 

Henceforth, we will adopt Equation~\eqref{N4} as the constitutive relation of NLG theory. 

In a previous attempt at finding nontrivial exact solutions of NLG, conformally flat spacetimes were considered~\cite{Bini:2016phe}. These are discussed in this paper in Appendices A and B. We work out the explicit form of the new constitutive relation for conformally flat spacetimes in Appendix A; however, we are still unable to solve the field equation of NLG. Instead, we discuss in the rest of this paper the issue of  whether de Sitter spacetime is a solution of NLG.

\section{Is de Sitter Spacetime a Solution of NLG?}

The spacetime of constant positive curvature is given by de Sitter metric, which is conformally flat and has the same form as Equation~\eqref{A1}, namely,   
\beq\label{S1}
ds^2 = \frac{1}{(\lambda\, t)^2}\,\eta_{\mu \nu} dx^\mu dx^\nu\,, \qquad \lambda :=  \left(\frac{\Lambda}{3}\right)^{1/2}\,.
\eeq
This is a solution of the vacuum Einstein equation with cosmological constant $\Lambda >0$; hence, it is  also a solution of NLG provided $\mathcal{N}_{\mu \nu} = Q_{\mu \nu}$. 

We can use the results of Appendix A in our calculations keeping in mind that $\lambda\, t = e^{-U}$. The new constitutive relation~\eqref{N4} is given by Equation~\eqref{A7}.  Let us write it in the form
\beq\label{S2}
N^{\mu \nu}{}_{\rho}(x) = 2\,\lambda^2 t\,(\eta^{0 \mu} \delta^\nu_\rho - \eta^{0 \nu} \delta^{\mu}_{\rho}) \mathcal{I}\,,
\eeq
where $U_\mu = -(1/t) \delta^0_\mu$ and $\mathcal{I}$ is the spacetime invariant defined by
\beq\label{S3}
\mathcal{I} :=  \int \mathcal{K}_{dS}(x, x')\,\sqrt{-g(x')} \,d^4x' =   \frac{1}{\lambda^4}\,\int \frac{1}{t'^4}\mathcal{K}_{dS}(x, x')\, d^4x'\,.
\eeq
We find that 
\beq\label{S4}
N^{0 i}{}_{\rho}(x) = - 2\,\lambda^2 t\, \delta^i_\rho\, \mathcal{I}\,,\qquad N^{i j}{}_{\rho}(x) = 0\,, \qquad N^{\alpha \beta}{}_{\beta} = - 6\,\lambda^2 t\, \delta^\alpha_0\,\mathcal{I}\,.
\eeq

On the other hand, we have from Equation~\eqref{A9},
\begin{equation}\label{S5}
Q_{\mu}{}^{\nu}=U_\mu\,N^{\nu \rho}{}_\rho - U_\rho\,N^{\nu \rho}{}_\mu-\frac{1}{2}\,\delta_{\mu}^{\nu}\,U_\alpha\,N^{\alpha \beta}{}_{\beta}\,.
\end{equation}
Therefore, 
\begin{equation}\label{S6}
Q_{\mu}{}^{\nu} = \lambda^2(4 \delta_\mu^0 \delta_0^\nu - \delta_{\mu}^{\nu})\mathcal{I}\,.
\end{equation}
This should equal $\mathcal{N}_{\mu}{}^{\nu}$, which can be expressed using Equation~\eqref{A8} as
\begin{equation}\label{S7}
\mathcal{N}_{\mu}{}^{\nu} (x) = t^3\,\frac{\partial}{\partial x^\beta}\,\Big(\frac{1}{t^3}\,N^{\nu \beta}{}_{\mu}\Big)\,.
\end{equation}
It is clear from Equation~\eqref{S4} that $\mathcal{N}_{0}{}^{0} = 0$, while $Q_{0}{}^{0} = 3 \lambda^2\,\mathcal{I}$. Indeed, it turns out that de Sitter spacetime is a solution of NLG provided
\begin{equation}\label{S8}
\mathcal{I} = 0\,.
\end{equation}

Let us next consider the coordinate transformation $(t, \mathbf{x}) \mapsto (T, \mathbf{X})$, where 
\begin{equation}\label{S9}
t = \frac{1}{\lambda} e^{-\lambda T}\,, \qquad  \mathbf{x} = \mathbf{X}\,.
\end{equation}
Then, de Sitter metric takes the form
\begin{equation}\label{S10}
ds^2 = -dT^2 + e^{2\lambda T} \delta_{ij}dX^i dX^j\,,
\end{equation}
from which we recover the Minkowski metric for $\lambda \to 0^{+}$. The fundamental tetrad frame is now given by $^{dS}e^{\mu}{}_{\hat 0} = \delta^\mu_0$ and $^{dS}e^{\mu}{}_{\hat i} = e^{-\lambda T}\delta^\mu_i$.

In these new coordinates, the vanishing invariant $\mathcal{I}$ can be written as 
\begin{equation}\label{S11}
\mathcal{I} =  \int \mathcal{K}_{dS}(X, X')\,e^{2\lambda T'} \,d^4X' = 0\,.
\end{equation}
In the limiting situation when $\lambda \to 0^{+}$, we expect $\mathcal{K}_{dS} \to \mathcal{K}_M$, where $\mathcal{K}_M$ is the causal constitutive kernel in the case of Minkowski spacetime. It therefore seems that in the limiting case of vanishing cosmological constant we must have 
\begin{equation}\label{S12}
\int \mathcal{K}_M(X, X') \,d^4X' = 0\,.
\end{equation}
However, according to Equation (7.151) of Ref.~\cite{BMB},  we have in NLG
\begin{equation}\label{S13}
\int \mathcal{K}_M(X, X') \,d^4X' = \hat{\chi}(0)\,,
\end{equation}
where $\hat{\chi}(0)$ is a negative number such that 
\begin{equation}\label{S14}
-1 < \hat{\chi}(0) < 0\,. 
\end{equation}
We conclude that de Sitter spacetime is \emph{not} a solution of NLG theory. The rest of this section is devoted to the calculation  of  $\mathcal{K}_{dS}$.

\subsection{$\mathcal{K}_{dS}$}

Let us first recall that in a global inertial frame with coordinates $X^\mu = ( T, \mathbf{X})$ in Minkowski spacetime, the fundamental inertial observers have adapted tetrads 
$^{M}e^{\mu}{}_{\hat \alpha} = \delta^\mu_\alpha$ and the world function is~\cite{BMB, Bini:2016phe}
\begin{equation}\label{W1}
^{M}\Omega(X, X') = -\frac{1}{2}\, [(T-T')^2 - |\mathbf{X} - \mathbf{X}'|^2]\,.
\end{equation}
The nonlocal causal kernel is a function of the invariants~\cite{BMB, Bini:2016phe}
\begin{equation}\label{W2}
^{M}\Omega_\mu(X, X')\,^{M}e^{\mu}{}_{\hat \alpha}|_{X} = -^{M}\Omega_\mu(X, X')\,^{M}e^{\mu}{}_{\hat \alpha}|_{X'} = \eta_{\alpha \beta}(X^\alpha - X'^\alpha)\,.
\end{equation}
As described in detail in Section (7.5) of~\cite{BMB}, the kernel of NLG for Minkowski spacetime can be written as
\begin{equation}\label{W3}
\mathcal{K}_M(X, X') = \Theta (T-T' - |\mathbf{X} - \mathbf{X}'|)\, k_M ( T-T', \mathbf{X} - \mathbf{X}')\,.
\end{equation}
Here, $\Theta$ is the Heaviside unit step function such that $\Theta(t) = 0$ for $t < 0$ and $\Theta(t) = 1$ for $t \ge 0$. Moreover, $k_M$ is a definite universal function as described in detail in Section (7.5) of~\cite{BMB}. This kernel is determined via its reciprocal kernel described briefly in Appendix C.  

The world function in de Sitter spacetime~\eqref{S1} is given by $^{dS}\Omega(x, x') = -\tau^2/2$, where~\cite{Ru}  
\begin{equation}\label{W4}
\cosh{(\lambda\,\tau)}=\frac{t^2+t'^2-|\mathbf{x}-\mathbf{x'}|^2}{2t\,t'} := \mathbb{Q}\,.
\end{equation}
Here,  $\mathbb{Q}\ge 1$ and  
\begin{equation}\label{W5}
\lambda\,\tau=\ln{(\mathbb{Q}+\sqrt{\mathbb{Q}^2-1})}\,.
\end{equation}
Under the coordinate transformation~\eqref{S9}, invariant $\mathbb{Q}$ takes the form
\begin{equation}\label{W6}
\mathbb{Q} = \cosh [\lambda(T-T')]- \mathbb{L}\,,
\end{equation}
where
\begin{equation}\label{W7}
\mathbb{L} := \frac{1}{2}\lambda^2\, e^{\lambda(T+T')}\,|\mathbf{X} - \mathbf{X}'|^2\,.
\end{equation}
One can check that in terms of the new coordinates, $^{dS}\Omega \to~ ^M\Omega$ as $\lambda \to 0$. Moreover, the requirement that $\mathbb{Q} \ge 1$ becomes $\cosh [\lambda(T-T')] \ge 1+\mathbb{L}$.  It is now straightforward to show that 
\begin{equation}\label{W8}
^{dS}\Omega_\mu(X, X')\,^{dS}e^{\mu}{}_{\hat 0}|_{X} = \frac{\lambda \tau}{\lambda \sinh (\lambda \tau)}\, \left\{-\sinh [\lambda(T-T')] +\mathbb{L}\right\}\,
\end{equation}
and
\begin{equation}\label{W9}
^{dS}\Omega_\mu(X, X')\,^{dS}e^{\mu}{}_{\hat i}|_{X} = \frac{\lambda \tau}{\sinh (\lambda \tau)}\,e^{\lambda T'} (\mathbf{X} - \mathbf{X}')\,.
\end{equation}
Similarly, 
\begin{equation}\label{W10}
^{dS}\Omega_\mu(X, X')\,^{dS}e^{\mu}{}_{\hat 0}|_{X'} = \frac{\lambda \tau}{\lambda \sinh (\lambda \tau)}\, \left\{\sinh [\lambda(T-T')] +\mathbb{L}\right\}\,
\end{equation}
and
\begin{equation}\label{W11}
^{dS}\Omega_\mu(X, X')\,^{dS}e^{\mu}{}_{\hat i}|_{X'} = -\frac{\lambda \tau}{\sinh (\lambda \tau)}\,e^{\lambda T} (\mathbf{X} - \mathbf{X}')\,.
\end{equation}
We note that $\mathcal{K}_{dS} (X, X')$ must be a function of the above invariants, which reduce to the corresponding Minkowski invariants when $\lambda \to 0$.  Based on these results, let 
\begin{equation}\label{W12}
\mathbb{U} := \frac{1}{\lambda}\,\ln{(\mathbb{S}+\sqrt{\mathbb{S}^2-1})}\,, \qquad \mathbb{S} := 1 + \mathbb{L}\,,
\end{equation}
\begin{equation}\label{W13}
\mathbb{V} := \frac{\lambda \tau}{\lambda \sinh (\lambda \tau)}\, \left\{\sinh [\lambda(T-T')] +\mathbb{L}\right\}\,,
\end{equation}
\begin{equation}\label{W14}
\mathbb{W} := \frac{\lambda \tau}{\sinh (\lambda \tau)}\,e^{\lambda T} (\mathbf{X} - \mathbf{X}')\,.
\end{equation}
Then, 
\begin{equation}\label{W15}
\mathcal{K}_{dS}(X, X') = \Theta (T-T' - \mathbb{U})\, k_M (\mathbb{V}, \mathbb{W})\,.
\end{equation}
For $\lambda \to 0$, we find $\mathcal{K}_{dS} \to \mathcal{K}_{M}$.

\section{Discussion}
Nonlocal gravity (NLG) has been patterned after the electrodynamics of media and therefore involves a nonlocal constitutive relation. Linearized NLG has been mainly studied thus far, since no exact nontrivial solution of the theory is known at present. This could be in part due to the complicated nature of the constitutive relation. In this paper, we adopt a new constitutive relation for NLG and use it to show that de Sitter spacetime is not a solution of NLG. The new constitutive relation, which is the same as the one recently suggested in Ref.~\cite{Puetzfeld:2019wwo}, is more physically motivated and simpler than the old one; besides, it appears to be more amenable to finding exact solutions of NLG.

\section*{ACKNOWLEDGMENTS}

I am grateful to Friedrich Hehl and Yuri Obukhov for valuable discussions. 

\appendix

\section{New Constitutive Relation for Conformally Flat Spacetimes}

Consider a conformally flat spacetime with
\begin{equation}\label{A1}
ds^2= e^{2U} \, \eta_{\mu \nu}\, dx^\mu\,dx^\nu\,,
\end{equation}
where $U(x)$ is a scalar function. The fundamental observers in this case are all at rest in space with adapted orthonormal tetrads 
\begin{equation}\label{A2}
e^\mu{}_{\hat{\alpha}}(x) = e^{-U}\,\delta^\mu_{\alpha}\,, \qquad     e_\mu{}^{\hat{\alpha}}(x) = e^{U}\,\delta_\mu^{\alpha}\,.
\end{equation}
The Einstein tensor for metric~\eqref{A1} is given by~\cite{Ste}
\begin{equation}\label{A3}
^0G_{\mu \nu}=-2\,(U_{\mu \nu}-U_\mu\,U_\nu)+\eta_{\mu \nu}\eta^{\alpha \beta}(U_\alpha\,U_\beta+2\,U_{\alpha \beta})\,.
\end{equation}
Here, $U_{\mu}:= \partial_{\mu}U$, $U_{\mu \nu} = \partial_{\mu}\partial_{\nu}U$, etc. The gravitational source could be a perfect fluid of energy density $\rho$ and pressure $p$ with energy-momentum tensor 
\begin{equation}\label{A4}
T_{\mu \nu}=\rho\,u_\mu\,u_\nu+p\,(g_{\mu \nu}+u_\mu\,u_\nu)\,,
\end{equation}
where $u^\mu$ is the fluid's 4-velocity vector. 

From the preferred tetrad frame field~\eqref{A2}, we find  the torsion tensor and the torsion vector
\begin{equation}\label{A5}
C_{\alpha \beta \gamma}=e^{2U}\,(U_{\alpha}\, \eta_{\beta \gamma}- U_{\beta}\, \eta_{\gamma \alpha})\,, \qquad  C_{\alpha}=-3\,U_{\alpha}\,. 
\end{equation}
Furthermore,  the contorsion tensor and the auxiliary torsion tensor are given by
\begin{equation}\label{A6}
 K_{\alpha \beta \gamma}=C_{\beta\gamma\alpha}\,, \qquad  \mathfrak{C}_{\alpha \beta \gamma}=-2\,C_{\alpha \beta \gamma}\,. 
\end{equation}
The torsion pseudovector vanishes in the conformally flat case, namely,  $\check{C}_\alpha=0$; hence, $X_{\mu \nu \rho}= \mathfrak{C}_{\mu \nu \rho}$. The new constitutive relation of NLG thus implies
\begin{equation}\label{A7}
N_{\mu \nu \rho}(x) = -2\, \int [U_{\mu}(x') \eta_{\nu \rho}\ - U_{\nu}(x') \eta_{\mu \rho}]\mathcal{K}(x, x')\,e^{3[U(x) + U(x')]}\, d^4x'\,.
\end{equation}

In terms of $N_{\mu \nu \rho}$,  we can write the nonlocal parts of the field equation~\eqref{G24} as 
\begin{equation}\label{A8}
\mathcal{N}_{\mu \nu}=e^{-U} \,\eta_{\nu \alpha}\,\frac{\partial}{\partial x^\beta}\,\Big(e^{3\,U}\,N^{\alpha \beta}{}_{\mu}\Big)\,,
\end{equation}
and 
\begin{equation}\label{A9}
Q_{\mu \nu}=U_\mu\,N_{\nu}{}^{\rho}{}_\rho-U_\rho\,N_{\nu}{}^{\rho}{}_\mu-\frac{1}{2}\,g_{\mu \nu}\,U_\alpha\,N^{\alpha \beta}{}_{\beta}\,.
\end{equation}
The symmetric parts of these tensors contribute to Equation~\eqref{G27}, while  the six supplementary constraints given by $\mathcal{N}_{[\mu \nu]} = Q_{[\mu \nu]}$ can be expressed as
\begin{equation}\label{A10}
g_{\mu \alpha}\,\frac{\partial N^{\alpha \beta}{}_{\nu}}{\partial x^\beta} - g_{\nu \alpha}\,\frac{\partial N^{\alpha \beta}{}_{\mu}}{\partial x^\beta} + U_\mu \,N_{\nu}{}^{\alpha}{}_{\alpha} - U_\nu \,N_{\mu}{}^{\alpha}{}_{\alpha} + 4 U_\beta\,( N_{\mu}{}^{\beta}{}_{\nu} - N_{\nu}{}^{\beta}{}_{\mu}) = 0\,.
\end{equation}

\section{2D Spacetimes}

Any two-dimensional spacetime is conformally flat~\cite{Ste}; therefore, its metric can be written as
\begin{equation}\label{B1}
ds^2=e^{2U(t,x)}(-dt^2+dx^2)\,.
\end{equation}
 In this case, $g_{\mu \nu} = \exp(2U)\,\tilde{\eta}_{\mu \nu}$, $\sqrt{-g}= \exp{(2U)}$ and $\tilde{\eta}_{\mu \nu} :=$ diag$(-1, 1)$. Using formulas given in Section (3.7) of Ref.~\cite{Ste}, one can show that the Ricci tensor, scalar curvature and Einstein tensor in 2D are given by
\begin{equation} \label{B2}
{}^0R_{\mu\nu}=-\tilde{\eta}_{\mu \nu}\,\tilde{\eta}^{\alpha \beta}U_{\alpha \beta}\,,\qquad
{}^0R=-2\,e^{-2U}\,\tilde{\eta}^{\alpha \beta}U_{\alpha \beta}\,,\qquad
{}^0G_{\mu\nu}=0\,.
\end{equation} 
 We assume that in a 2D spacetime, $\Lambda=0$ and $T_{\mu\nu}=0$; therefore, any 2D spacetime is a solution of the GR field equation. 
 
On the other hand, the field equation of  NLG  in 2D reduces to ${\mathcal N}_{\mu\nu}= Q_{\mu\nu}$, where both sides vanish if $N_{\alpha\beta \gamma} = 0$. Let us note that in 2D, 
\begin{equation} \label{B3}
C_{\mu\nu \rho}= U_\mu g_{\nu \rho} - U_\nu g_{\mu \rho}\,, \quad K_{\mu \nu \rho} = C_{\nu \rho \mu}\,, \quad C_\mu = -U_\mu\,, \quad \check{C}_\alpha=0\,
\end{equation} 
and
\begin{equation} \label{B4}
\mathfrak{C}_{\mu\nu \rho} = 0\,.
\end{equation}
The connection of $N_{\alpha \beta \gamma}$ and $\mathfrak{C}_{\mu\nu \rho}$ via the new (or old) constitutive relation of NLG then implies that $N_{\mu\nu \rho} = 0$ in 2D. This circumstance leads to the fact that, just as in GR, the field equation of NLG is always satisfied in 2D. 

In Section 5 of Ref.~\cite{Bini:2016phe}, the nonzero 4D value of $\mathfrak{C}_{\mu\nu \rho}$ was inadvertently employed in 2D leading to erroneous conclusions. Indeed, 
 the parts of the treatment given in Section 5 of Ref.~\cite{Bini:2016phe} that use $\mathfrak{C}_{\mu\nu \rho} \ne 0$ in 2D are incorrect.   

\section{Reciprocal kernel $\mathcal{R}_M(X-X')$} 

The kernel of NLG for Minkowski spacetime, $\mathcal{K}_M(X, X')$, or equivalently,  $\mathcal{K}_M(X-X')$, given by Equation~\eqref{W3}, has a reciprocal kernel $\mathcal{R}_M(X-X')$ such that 
\begin{equation}\label{C1}
\mathcal{K}_M(X-X') + \mathcal{R}_M(X-X') + \int \mathcal{K}_M(X-X'') \mathcal{R}_M(X''-X')\,d^4X'' = 0\,,
\end{equation}
see Section (7.2) of~\cite{BMB}. In principle, the kernel can be determined once its reciprocal is known. According to Equation (7.143) of~\cite{BMB}, 
\begin{equation}\label{C2}
\mathcal{R}_M(X-X') = \nu \,e^{-\nu(T-T' - |\mathbf{X} - \mathbf{X}'|)} \,\Theta (T-T' - |\mathbf{X} - \mathbf{X}'|)\,q(|\mathbf{X} - \mathbf{X}'|)\,,
\end{equation}
where $\nu^{-1}$ is a constant length to be determined via future observational data and $q(r)$ is a spherically symmetric $L^1$ and $L^2$ function of  $r = |\mathbf{X} - \mathbf{X'}|$. Two simple possible forms for $q$ have been considered based on the nearly flat rotation curves of nearby spiral galaxies, namely,  
\begin{equation}\label{C3}
 q_1 (r) = \frac{1}{4\pi \ell_0} \,\frac{1+\mu_0\, (a_0+r)}{r\,(a_0 + r)}\,e^{-\mu_0\,r}\,, \qquad q_2 (r) = \frac{r}{a_0 + r}\,q_1(r)\,.
\end{equation}  
Here, $\ell_0$, $\mu_0^{-1}$ and $a_0$ are three constant lengths; moreover, $\ell_0$ is the main nonlocality parameter, since the reciprocal kernel tends to zero as $\ell_0$ tends to infinity. Observational data regarding nearby spiral galaxies and clusters of galaxies are consistent with~\cite{Rahvar:2014yta}
\begin{equation}\label{C4}
 \mu_0 = 0.059 \pm 0.028~{\rm kpc}^{-1}\,, \qquad  \ell_0  \approx 3\pm 2~{\rm kpc}\,.
\end{equation}
Finally, solar system data provide a lower bound for $a_0$, namely, $a_0 \gtrsim 10^{15}$~cm~\cite{Chicone:2015coa, Roshan:2022ypk}.


\begin{thebibliography}{00}

\bibitem{BMB}
Mashhoon, B. 
\emph{Nonlocal Gravity};
Oxford University Press: Oxford, UK, 2017.


\bibitem{L+L}
Landau, L.D.; Lifshitz, E.M. 
\emph{Electrodynamics of Continuous Media}; Pergamon: Oxford, UK, 1960.


\bibitem{Jack}
Jackson, J.D. \emph{Classical Electrodynamics}, 3rd ed.; Wiley: Hoboken, NJ, USA, 1999. 


\bibitem{HeOb}  
Hehl, F.W.; Obukhov, Y.N.
\emph{Foundations of Classical Electrodynamics: Charge, Flux, and Metric}; 
Birkh\"auser: Boston, MA, USA, 2003. 

\bibitem{Einstein}
Einstein, A. 
\emph{The Meaning of Relativity}; 
Princeton University Press: Princeton, NJ, USA, 1955.

\bibitem{We}
Weitzenb\"ock, R. 
{\it Invariantentheorie};
Noordhoff: Groningen, the Netherlands, 1923.

\bibitem{Maluf:2011kf}
Maluf, J.W.; Faria, F.F.
Conformally invariant teleparallel theories of gravity.
\emph{Phys. Rev. D} {\bf 2012},  \emph{85}, 027502.
[arXiv:1110.3095 [gr-qc]]

\bibitem{Aldrovandi:2013wha}
Aldrovandi, R.; Pereira, J.G.
\emph{Teleparallel Gravity: An Introduction};
Springer: New York, 2013.

\bibitem{Maluf:2013gaa}
Maluf, J.W.
The teleparallel equivalent of general relativity.
\emph{Ann. Phys. (Berlin)} {\bf 2013}, \emph{525}, 339--357.
[arXiv:1303.3897 [gr-qc]]


\bibitem{Cho}
Cho, Y.M.
Einstein Lagrangian as the translational Yang-Mills Lagrangian.
\emph{Phys. Rev. D} {\bf 1976},  \emph{14}, 2521--2525.

\bibitem{Hehl:2008eu}
Hehl, F.W.; Mashhoon, B.
 Nonlocal Gravity Simulates Dark Matter.
\emph{Phys. Lett. B} \textbf{2009}, \emph{673}, 279--282.  
[arXiv:0812.1059 [gr-qc]]

              

\bibitem{Hehl:2009es}
Hehl, F.W.; Mashhoon, B.
 Formal framework for a nonlocal generalization of Einstein's theory of gravitation.
\emph{Phys. Rev. D} \textbf{2009}, \emph{79}, 064028.  
[arXiv:0902.0560 [gr-qc]]


\bibitem{BlHe}
Blagojevi\'c, M.; Hehl, F.W., Eds. 
\emph{Gauge Theories of Gravitation}; 
Imperial College Press: London, UK, 2013.

\bibitem{Itin:2018dru}
Itin, Y.; Obukhov, Y.N.; Boos, J.; Hehl, F.W.
Premetric teleparallel theory of gravity and its local and linear constitutive law.
\emph{Eur. Phys. J. C} \textbf{2018},  \emph{78}, 907.
[arXiv:1808.08048 [gr-qc]]

\bibitem{Mashhoon:2019jkq}
Mashhoon, B.; Hehl, F.W.
 Nonlocal Gravitomagnetism.
\emph{Universe} \textbf{2019}, \emph{5},  195.
[arXiv:1908.05431 [gr-qc]]


\bibitem{Hehl:2020hhp}
Hehl, F.W.; Obukhov, Y.N.
Conservation of Energy-Momentum of Matter as the Basis for the Gauge Theory of Gravitation.
\emph{Fundam. Theor. Phys.} \textbf{2020}, \emph{199}, 217--252.
[arXiv:1909.01791 [gr-qc]]



\bibitem{Puetzfeld:2020tkp}
Puetzfeld, D.; Obukhov, Y.N.
Generalized nonlocal gravity framework based on Poincar\'e gauge theory.
\emph{Phys. Rev. D} \textbf{2020}, \emph{101}, 104054.
[arXiv:2001.11073 [gr-qc]]

\bibitem{Obukhov:2020uan}
Obukhov, Y.N.; Hehl, F.W.
General relativity as a special case of Poincar\'e gauge gravity.
\emph{Phys. Rev. D} \textbf{2020}, \emph{102},  044058.
[arXiv:2007.00043 [gr-qc]]

\bibitem{Rahvar:2014yta}
Rahvar, S.; Mashhoon, B.
 Observational Tests of Nonlocal Gravity: Galaxy Rotation Curves and Clusters of Galaxies.
\emph{Phys. Rev. D} \textbf{2014}, \emph{89}, 104011.
[arXiv:1401.4819 [astro-ph.GA]]

\bibitem{Chicone:2015coa}
Chicone, C.; Mashhoon, B.
 Nonlocal Gravity in the Solar System.
\emph{Classical Quantum Gravity} \textbf{2016}, \emph{33},  075005.
[arXiv:1508.01508 [gr-qc]]



\bibitem{Roshan:2021ljs}
Roshan, M.; Mashhoon, B.
 Dynamical Friction in Nonlocal Gravity.
\emph{Astrophys. J.} {\bf 2021}, \emph{922},  9.
[arXiv:2107.05841 [gr-qc]]


\bibitem{Roshan:2022zov}
Roshan, M.; Mashhoon, B.
 Characteristics of Effective Dark Matter in Nonlocal Gravity.
\emph{Astrophys. J.} \textbf{2022}, \emph{934},  9.
[arXiv:2201.12852 [astro-ph.GA]]

\bibitem{Roshan:2022ypk}
Roshan, M.; Mashhoon, B.
Nonlocal Gravity: Modification of Newtonian Gravitational Force in the Solar System.
\emph{Universe} \textbf{2022}, \emph{8}, 470.
[arXiv:2205.13276 [gr-qc]]


\bibitem{Sy}
Synge, J.L.
\emph{Relativity: The General Theory}; 
North-Holland: Amsterdam, the Netherlands, 1971.



\bibitem{Puetzfeld:2019wwo} 
  Puetzfeld, D.; Obukhov, Y.N.; Hehl, F.W.
  Constitutive law of nonlocal gravity.
  \emph{Phys. Rev. D} {\bf 2019}, \emph{99}, no. 10, 104013. 
  [arXiv:1903.04023 [gr-qc]]
  
\bibitem{Bini:2016phe}
Bini, D.; Mashhoon, B.
 Nonlocal gravity: Conformally flat spacetimes.
\emph{Int. J. Geom. Meth. Mod. Phys.} \textbf{2016}, \emph{13},  1650081.
[arXiv:1603.09477 [gr-qc]]

  
\bibitem{Ru}
Ruse, H.S.
The Potential of an Electron in a Space-Time of Constant Curvature.
\emph{Quart. J. Math. (Oxford)} \textbf{1930}, \emph{1},  146--155.  


\bibitem{Ste} 
Stephani, H.; Kramer, D.; MacCallum, M.A.H.; Hoenselaers, C.; Herlt, E.
\emph{Exact Solutions of Einstein's Field Equations}, 2nd ed.;
Cambridge University Press: Cambridge, UK, 2003.




\end{thebibliography}
\end{document}